\documentclass[apl,aps,nofootinbib,notitlepage,superscriptaddress,twocolumn,prd,10pt]{revtex4-1}

\usepackage{amsmath, amsfonts,bm,amssymb}
\usepackage{subfig}
\usepackage[dvipsnames]{xcolor}
\usepackage{graphicx}
\usepackage{soul}
\definecolor{myred}{rgb}{0.9, 0.17, 0.31}
\usepackage[colorlinks=true,hyperfootnotes=true, linkcolor=myred, citecolor=cyan, urlcolor=magenta]{hyperref}

\def\dd{\mathrm{d}}
\newcommand{\be}{\begin{equation}} 
\newcommand{\ee}{\end{equation}}
\newcommand{\ba}{\begin{eqnarray}}
\newcommand{\ea}{\end{eqnarray}}

\begin{document}

\title{Vector wormholes as conduits for matter interaction}

\author{Nailya Ganiyeva}
\email{fc57452@alunos.fc.ul.pt}
\affiliation{Instituto de Astrof\'isica e Ci\^encias do Espa\c{c}o,\\ 
Faculdade de Ci\^encias da Universidade de Lisboa,  \\ Campo Grande, PT1749-016 
Lisboa, Portugal}

\author{Bruno J. Barros}
\email{bjbarros@fc.ul.pt}
\affiliation{Instituto de Astrof\'isica e Ci\^encias do Espa\c{c}o,\\ 
Faculdade de Ci\^encias da Universidade de Lisboa,  \\ Campo Grande, PT1749-016 
Lisboa, Portugal}

\author{\'Alvaro de la Cruz-Dombriz}
\email{alvaro.dombriz@usal.es}
\affiliation{Cosmology and Gravity Group, Department of Mathematics and Applied Mathematics,
University of Cape Town, Rondebosch 7700, Cape Town, South Africa}
\affiliation{ Departamento de F\'isica Fundamental, Universidad de Salamanca,
    P. de la Merced, 37008 Salamanca, Spain}

\author{Francisco S.N. Lobo}
\email{fslobo@fc.ul.pt}
\affiliation{Instituto de Astrof\'isica e Ci\^encias do Espa\c{c}o,\\ 
Faculdade de Ci\^encias da Universidade de Lisboa,  \\ Campo Grande, PT1749-016 
Lisboa, Portugal}
\affiliation{Departamento de F\'{i}sica, Faculdade de Ci\^{e}ncias, Universidade de Lisboa, Edifício C8, Campo Grande, PT1749-016 Lisbon, Portugal}

\date{\today}

\begin{abstract}

In this work, we focus on the dynamics of a massive one-form field, \textbf{B}, often referred to simply as a vector field, that is minimally coupled to standard Einstein gravity. In the framework of four-dimensional spacetimes, the theory of a massive one-form propagates three massive vector degrees of freedom. The inclusion of a self-interacting potential in this theory results in the breaking of gauge invariance. The breaking of such a fundamental symmetry in Classical Electromagnetism may introduce a ghost mode in massive vector theories, which generally leads to their instability.
However, in the context of wormhole physics, the existence of at least one ghost degree of freedom turns out to be a necessary condition to support these exotic geometries within effective field theories. This requirement serves as a strong motivation for our work, wherein we explore the role and phenomenology of massive one-forms, minimally coupled to Einstein gravity, in providing the necessary conditions to sustain wormhole spacetimes.
We further analyze the coupling of matter fields to such a vector field through conformal couplings and explore their impact on energy conditions and the physical viability of wormhole solutions.

\end{abstract}

\maketitle

\section{Introduction}\label{sec:intro}

A wormhole is a topological structure that connects either two distinct spacetime manifolds or two different regions within the same spacetime manifold. These objects have been extensively studied in the context of General Relativity (GR) \cite{Morris:1988cz, Morris:1988tu, Visser:1995cc, Lemos:2003jb, Visser:2003yf, Kar:1995ss,Lobo:2017cay}, where they exhibit a crucial limitation: the requirement for a wormhole spacetime to be traversable implies the so-called flaring-out condition \cite{Morris:1988cz}, which, through the Einstein field equations, leads to the violation of the null energy condition (NEC) and, by extension, all the other energy conditions \cite{Visser:1995cc,Hawking:1973uf}. 
In fact, classical forms of matter are generally understood to satisfy these standard energy conditions \cite{Hawking:1973uf}, whereas matter distribution that violates the NEC is termed as \textit{exotic} matter.
However, certain quantum fields, such as in the Casimir effect and Hawking radiation, amongst others, are known to violate these conditions~\cite{Visser:1995cc}. In the context of quantum systems within classical gravitational backgrounds, violations of the null or weak energy conditions are typically constrained to small magnitudes 
\cite{Tipler:1978zz, Ford:1994bj}.

The transient nature of negative energy states, which are balanced by subsequent positive energy states, ensures an overall balance with conservation laws~\cite{Ford:1995wg,Ford:1999qv}. This behavior motivated the formulation of constraints on negative energy densities for free, quantized, massless scalar fields within quantum field theory, culminating in the development of the Quantum Inequality (QI)~\cite{Ford:1995wg}. The QI imposes strict bounds on both the magnitude and duration of negative energy violations, providing critical insights into their spatial and temporal distribution.
In wormhole physics, the QI plays an important role by restricting the extent to which negative energy, required to sustain traversable wormholes, can exist \cite{Ford:1995wg}. These limits suggest significant challenges for constructing and maintaining wormholes under realistic quantum field conditions, as the necessary exotic matter must comply with these stringent constraints.

To overcome the limitations of GR regarding the requirement of exotic matter to sustain traversable wormholes, extended theories of gravity beyond the Einsteinian paradigm provide a compelling alternative framework~\cite{Agnese:1995kd,Nandi:1997en,LaCamera:2003zd,Garattini:2007ff,Lobo:2007zb,Lobo:2008zu,Garattini:2008xz,Lobo:2010sb,MontelongoGarcia:2011ag,Garattini:2011fs}. These theories introduce either additional terms or interactions in the gravitational sector capable of generating the necessary conditions for maintaining a traversable wormhole throat, while ensuring that the matter sources involved do not violate the energy conditions and remain non-exotic \cite{Harko:2013yb}.
Several classes of modified gravity theories have been extensively studied for their potential to support wormhole geometries. Amongst them, $f(R)$ gravity and its generalizations stand out as prominent candidates. By modifying the functional form of the Ricci scalar $R$ in the Einstein-Hilbert action, these theories provide the flexibility to achieve the desired curvature effects without resorting to exotic matter~\cite{Lobo:2009ip,Capozziello:2012hr,Rosa:2018jwp,Rosa:2021yym,Rosa:2023olc,Rosa:2022osy,Rosa:2023guo}. Extensions of $f(R)$ gravity often include additional degrees of freedom or higher-order curvature invariants, further expanding their applicability to wormhole physics.

Theories incorporating curvature-matter couplings represent another significant avenue of research. Therein, matter and curvature are coupled in a nonminimal manner, leading to effective energy-momentum tensors that can mimic exotic matter-like effects without requiring unphysical matter distributions~\cite{Garcia:2010xb,MontelongoGarcia:2010xd,Ayuso:2014jda,Alvarenga:2013syu}.
Similarly, theories introducing additional fundamental fields, such as scalar, vector, or tensor fields, provide mechanisms for modifying gravitational interactions, enabling the construction of wormhole solutions consistent with non-exotic matter~\cite{Harko:2013yb,Anchordoqui:1996jh}.
Einstein-Cartan gravity, which extends GR by incorporating torsion as an additional geometric property of spacetime, has also been shown to allow for traversable wormhole solutions. As a matter of fact, the inclusion of torsion creates modifications in the field equations that can generate effective energy conditions favorable for wormhole stability~\cite{DiGrezia:2017daq}.

In the realm of higher-dimensional and string-inspired theories, Gauss-Bonnet gravity and its extensions play a crucial role. Indeed, by adding the Gauss-Bonnet term, i.e. a higher-order curvature invariant, to the action, these theories provide corrections to GR that are particularly relevant in higher-dimensional spacetimes. These corrections have been demonstrated to support traversable wormholes with matter distributions that satisfy standard energy conditions~\cite{Bhawal:1992sz,Dotti:2007az,Mehdizadeh:2015jra}.
Finally, braneworld scenarios, which arise from higher-dimensional theories like those inspired by string theory, offer another promising framework for wormhole physics. In these models, our observable universe is a four-dimensional hypersurface (brane) embedded in a higher-dimensional bulk. Thus, the modified gravitational dynamics on the brane, induced by the bulk geometry, can support wormhole solutions without invoking exotic matter~\cite{Bronnikov:2002rn,Lobo:2007qi}.
All these approaches highlight the versatility of modified gravity theories in addressing the challenges posed by GR in wormhole physics. By providing mechanisms to circumvent the exotic matter requirement, they open up new possibilities for the theoretical study of traversable wormholes.

One way to overcome the use of exotic matter, is to consider wormhole solutions supported by a massive vector fields. By including an extra field — specifically a vector field — it becomes possible to concentrate the necessary exotic features within a distinct part of the system. This approach enables the remaining matter content to behave in a conventional and physically reasonable manner. Structuring the model this way is theoretically attractive, as it allows for scenarios where the violation of energy conditions is confined to a sector that could plausibly originate from broader theoretical frameworks, such as beyond-standard-model physics or modifications to general relativity. In this context, we explore the possibility that wormholes can be supported by $n$-forms. In fact, forms are a common feature of string theories and naturally appear in low-energy effective actions. Antisymmetric gravity is one example, stemming from Einstein's attempt to unify GR and Electromagnetism resorting to an asymmetric metric. While this approach introduced Maxwell-like terms, it failed to accurately describe the Lorentz force. Nonetheless, nonsymmetric extensions of GR remain significant, particularly for their theoretical and phenomenological roles, such as in dark matter alternatives involving antisymmetric tensor fields \cite{Moffat:1994hv,Prokopec:2006kr}. These fields also provide a mechanism for generating propagating torsion. However, breaking gauge invariance in actions endowed with forms may trigger ghost or tachyon instabilities. Despite these challenges, nonlinear dynamics can sometimes suppress these instabilities, and stable formulations have been developed for certain cases \cite{Damour:1992bt,Clayton:1996dz}.

The study of phenomenology of vector fields and their generalizations to forms \cite{Tasinato:2013oja}, and applications to cosmology \cite{Esposito-Farese:2009wbc,Ngampitipan:2011se,Bartolo:2012sd,Mulryne:2012ax,Koivisto:2012xm,Kumar:2014oka} and astrophysics \cite{Barros:2018lca,Barros:2020ghz,Bouhmadi-Lopez:2020wve,Barros:2021jbt,Bouhmadi-Lopez:2021zwt,Tangphati:2023uxt,Barros:2023pre} has been explored extensively in the literature. In particular, the cosmology of self-interacting three-forms was investigated in \cite{Koivisto:2009ew}, where it was shown that the minimally coupled canonical theory can naturally produce a diverse range of isotropic background dynamics, including scaling solutions, transient acceleration, and phantom crossing. Moreover, it has been shown that three-form fields can lead to viable cosmological scenarios for both inflation and dark energy, potentially yielding observable signatures that distinguish them from standard single scalar field models. \cite{Koivisto:2009fb,Germani:2009gg}. 

In this work, we extend previous studies in wormholes physics \cite{Barros:2018lca,Bouhmadi-Lopez:2021zwt,Tangphati:2023uxt}, by investigating the role of a massive one-form field, minimally coupled to Einstein gravity, in sustaining wormholes. In four-dimensional spacetime, these fields propagate three massive vector degrees of freedom, and the presence of a self-interacting potential breaks gauge invariance, possibly introducing a ghost mode that typically causes an instability. However, such ghost modes are essential to support wormhole geometries, motivating this study. Additionally, in this context we examine matter-field couplings through conformal interactions to assess their influence on the energy conditions and the viability of wormhole solutions.

This manuscript is organized as follows: In Sec. \ref{sec:theory0}, we present the action and the field equations for a massive one-form field, minimally coupled to Einstein gravity, when embedded in a wormhole geometry. Next, in Sec. \ref{sec:numerics}, we explore specific wormhole solutions and analyze the energy conditions. In Sec. \ref{sec:couple}, we study matter-field couplings through conformal interactions to assess their impact on the energy conditions and the viability of wormhole solutions. Finally, in Sec. \ref{sec:conclusions}, we summarize our results and conclusions. Unless otherwise stated, in the following we work in ${c=8\pi G =1}$ units.

\section{Theory and framework}\label{sec:theory0}

\subsection{Action}\label{sec:theory1}

Let $\left(\mathcal{M},\textbf{g}\right)$ be a smooth Riemannian space, with $\mathcal{M}$ here representing the spacetime manifold and $\textbf{g}$ a metric tensor. A one-form, $\textbf{B}:T^*\mathcal{M}\rightarrow \mathbb{R}$, inhabits smooth sections of the cotangent bundle of $\mathcal{M}$. In this work, we focus on a massive one-form, \textbf{B}, sometimes referred to simply as a vector, minimally coupled to standard Einstein gravity, described by the following action \cite{Koivisto:2009sd}
\ba\label{eq:action}
\mathcal{S}=\int \dd^4x\sqrt{-g}\left[ \frac{R}{2}  -\frac{1}{4}F_{\mu\nu}F^{\mu\nu}-V\left(B_\mu B^\mu\right) \right]
	\nonumber \\
+\,\mathcal{S}_m\left(\psi,g_{\mu\nu}\right)\,,
\ea
where 
$g$ denotes the determinant of the metric $\textbf{g}$, $R$ is the curvature scalar $R=g^{\mu \nu}R_{\mu \nu}$, with $R_{\mu \nu}$ the corresponding Ricci tensor, $V$ is the self-interacting potential, $\mathcal{S}_m$ denotes the action for the ordinary matter fields, $\psi$. ${\textbf{F}=\dd\textbf{B}}$ is the strength tensor of the one-form, and possesses the components,
\be\label{eq:strength}
F_{\mu\nu}=2\nabla_{[\mu}B_{\nu ]} \equiv \partial_\mu B_\nu - \partial_\nu B_\mu\,,
\ee
which by definition is a closed form, i.e., ${\dd\textbf{F}=0}$. 

Maxwell's theory is trivially reproduced with $V=0$. While the standard Proca action has ${V\propto m^2B_\mu B^\mu}$ \cite{Itzykson:1980rh}, the action \eqref{eq:action} can be framed with the work in Ref.~\cite{Heisenberg:2014rta}, where the full general action for a massive vector field was considered (see also Ref.~\cite{BeltranJimenez:2016afo}). The massive one-form theory in 4-dimensions as per \eqref{eq:action} propagates three massive vector degrees of freedom \cite{Germani:2009iq}. The presence of a self interacting potential in the total action breaks gauge invariance, present in classical Electromagnetism, under ${B_{\mu}\rightarrow B_{\mu}+\partial_\mu\lambda}$. It is known \cite{Esposito-Farese:2009wbc} that the breaking of this symmetry may introduce a ghost mode in massive vector theories, rendering them unstable. However, for wormhole physics, it is known that to support the throat of a wormhole at least one ghost degree of freedom is required \cite{Bronnikov:2006pt}. For simplicity let us introduce the notation where to square a tensor denotes contraction of all the indices, i.e.,
\be\label{eq:squares}
F^2 \equiv F_{\mu\nu}F^{\mu\nu}\quad\quad\text{and}\quad\quad B^2\equiv B_\mu B^{\mu}\,.
\ee

As usual, the energy-momentum tensor of the $i$-th species, $T^{(i)}_{\mu \nu}$, is defined through the variation of the Lagrangian $\mathcal{L}_i$ with respect to the metric $g_{\mu\nu}$ as
\begin{equation}\label{eq:def_t}
T^{(i)}_{\mu\nu}=-\frac{2}{\sqrt{-g}}\frac{\delta\left(\sqrt{-g}\mathcal L_i\right)}{\delta g^{\mu\nu}}.
\end{equation}
Thus, identifying in Eq.~\eqref{eq:action}, the vector field Lagrangian as,
\be
\mathcal{L}_B=-\frac{1}{4}F_{\mu\nu}F^{\mu\nu}-V\left(B_\mu B^\mu\right)\,,
\ee
the energy-momentum tensor of the field becomes
\be\label{eq:emtensor}
T^{(B)}_{\mu\nu}=F_{\mu}{}^{\alpha}F_{\nu\alpha}+2\frac{\partial V}{\partial B^2}B_\mu B_\nu -g_{\mu\nu}\left[ \frac{F^2}{4}+V\left(B^2\right) \right]\,.
\ee

From Eq. \eqref{eq:action} we obtain the equations of motion for \textbf{B}
\be\label{eq:eom_general}
\nabla_\alpha F^{\alpha\mu}=2\frac{\partial V}{\partial B^2}B^\mu\,,
\ee
where, due to antisymmetry of the strength tensor $F_{\mu\nu}$ we have the additional constraint
\be\label{eq:constraint}
\nabla\cdot\left(\frac{\partial V}{\partial B^2}\textbf{B}\right)=0\,.
\ee

\subsection{Wormhole metric and matter distribution}

In this work we consider static and spherically symmetric traversable wormhole solutions, given by the following metric written in the usual Schwarzschild coordinates $\left(t,r,\theta,\phi\right)$ \cite{Morris:1988cz} 
\be\label{eq:metric}
\dd s^2 = -{\rm exp}\left(2\Phi(r)\right)\dd t^2 + \frac{\dd r^2}{1-b(r)/r}+r^2\dd\Omega^2\,,
\ee
with ${\Phi\left(r\right)}$ being the redshift function, ${b\left(r\right)}$ the shape function, and ${\dd\Omega^2=\dd\theta^2+\sin^2\theta \, \dd\phi^2}$ the solid angle surface element. For the wormhole to be traversable, the functions $\Phi(r)$ and $b(r)$ must satisfy certain requirements. First, the redshift function must remain finite throughout the spacetime, i.e., ${|\Phi(r)| < \infty}$, to avoid event horizons in the spacetime, thus allowing an observer to traverse through the wormhole's interior in both directions. Additionally, at the wormhole's throat ${r = r_0}$, a geometrical requirement known as the flaring-out condition, expressed as ${(b - b'r)/b^2 > 0}$ \cite{Morris:1988cz}, must be imposed. Such a condition is described by the following boundary conditions on the shape function
\begin{equation}\label{def_flaring}
b(r_0)=r_0, \qquad b'(r_0)<1 \,,
\end{equation}
with a prime denoting a derivative with respect to $r$. 

Endowing the spacetime with a local system of coordinates, here represented by $x^\mu$, allows us to express the form field in such a system, i.e.,
\be
\textbf{B}=B_\mu(x^\mu)\dd x^\mu\,.
\ee
We can thus choose a specific parametrization for our vector compatible with the symmetries of the metric \eqref{eq:metric}. One possible way to achieve this is through the ansatz
\be\label{eq:B_param}
B_\mu=\sqrt{-g_{tt}}\,\zeta(r)\delta^t_{\;\mu}\,,
\ee
where the scalar function $\zeta(r)$ fully describes the components of the field. This particular parametrization yields, for the nonzero components of the strength tensor Eq. \eqref{eq:strength}
\be
F_{rt}=-F_{tr}=\sqrt{-g_{tt}}\left(\zeta'+\zeta\Phi'\right)={\text{e}}^{\Phi}\left(\zeta'+\zeta\Phi'\right)\,.
\ee
We can now express all the quantities in terms of ${\zeta=\zeta(r)}$. The contractions in Eq.~\eqref{eq:squares} now become
\be\label{eq:squares2}
F^2 = -2\left(1-\frac{b}{r}\right)\left(\zeta'+\zeta\Phi'\right)^2\,, \qquad B^2=-\zeta^2\,,
\ee
respectively, where we have omitted the dependencies on the functions ${b\equiv b(r)}$ and ${\Phi \equiv \Phi(r)}$. 
It is to be noted that the constraint Eq.~\eqref{eq:constraint} is trivially satisfied at the background level.

The energy-momentum components in Eq. \eqref{eq:emtensor} define the energy density $\rho_\zeta$, radial tension $\tau_\zeta$, and pressure $p_\zeta$, of the vector field, given by
\ba
\rho_\zeta &=& -T^{(B)\,t}{}
_t = -\frac{F^2}{4}+V-\zeta V_{,\zeta}\,,\label{eq:rho} \\
\tau_\zeta &=& -T^{(B)\,r}{}
_r = -\frac{F^2}{4}+V\,,\label{eq:pr} \\
p_\zeta &=& T^{(B)\,\theta}{}
_\theta = T^{(B)\,\phi}{}
_\phi= -\frac{F^2}{4}-V\,,\label{eq:p}
\ea
respectively, where ${V_{,\zeta}=\partial V/\partial \zeta}$. 

As we later present in Sec.~\ref{sec:EC}, the condition for the vector field to support the wormhole geometry, violating the energy conditions, is that ${\rho_\zeta-\tau_\zeta<0}$. It is easy to see that this condition is intimately related with the presence of ghosts in the theory. In Sec.~II B of Ref.~\cite{Esposito-Farese:2009wbc} the authors derive the conditions for a general massive vector field theory to be free from instabilities. This model falls within such case. Through an Hamiltonian stability analysis, the condition for the Hamiltonian for this present theory to be bounded from below, rendering the theory ghost-free, is ${\dd V/\dd B^2>0}$. Through Eqs.~\eqref{eq:rho} and \eqref{eq:pr}, we have that
\be
\rho_\zeta-\tau_\zeta = -\zeta \frac{\dd V}{\dd\zeta} = \zeta^2\frac{\dd V}{\dd B^2}\,,
\ee
which, in order for the wormhole to be supported by the vector field, must be negative. This implies that ${V'(B^2)<0}$, which is inconsistent with the condition required for the theory to remain free of ghost instabilities. Thus, we conclude that in order for the wormhole to be supported by the vector field, there will be at least one ghost mode in the theory.

The dynamics of ${\zeta}$ can be found evaluating the ${\mu=t}$ component of Eq.~\eqref{eq:eom_general} on our background \eqref{eq:metric}, %
\begin{eqnarray}
\label{eq:eom_zeta}
&&\left(\zeta''+\zeta'\Phi'+\zeta\Phi''\right)\left(1-\frac{b}{r}\right)
	\nonumber \\
&&+\,\frac{1}{r}\left(\zeta'+\zeta\Phi' \right)\left( 2-\frac{3}{2}\frac{b}{r}-\frac{b'}{2} \right)+V_{,\zeta}=0\,.
\end{eqnarray}

On the other hand, the non-vanishing components of the geometrical counterpart of the field equations, ${G_{\mu\nu}\equiv R_{\mu\nu}-\frac{1}{2}Rg_{\mu\nu}}$,  yield
\begin{eqnarray}
G^t{}_t &=& -\frac{b'}{r^2}\,, \label{ee1} 	\\
G^r{}_r &=&- \frac{b}{r^3} + 2\left( 1-\frac{b}{r} \right)\frac{\Phi'}{r}\,, \label{ee2} \\
G^\theta{}_\theta &=& G^\phi{}_\phi = \frac{b-b'r}{2r^3}+\left( \Phi'' +\Phi'^2 \right)\left(1-\frac{b}{r}\right)
	\nonumber \\
&& \hspace{1cm} +\,\frac{\Phi'}{2r}\left(2-\frac{b}{r}-b'\right)\,. \label{ee3}
\end{eqnarray}

Regarding the ordinary matter sector, we assume that the distribution of matter is described by an anisotropic perfect fluid, for which the corresponding energy-momentum tensor $T^{(m)}_{\mu\nu}$ takes the following form
\be
T^{(m)}_{\mu\nu} = (\rho_m+p_m)u_{\mu}u_{\nu}+p_mg_{\mu\nu} - (\tau_m+p_m)\chi_{\mu}\chi_{\nu}\,,
\label{EM-matter}
\ee
with $u_{\mu}$ being the four-velocity vector and ${\chi^{\mu} = \delta^{\mu}_r \sqrt{1-b/r}}$ the radial component of a spacelike unit vector. Thus the diagonal of the energy-momentum tensor gives the energy density ${\rho_m\equiv\rho_m\left(r\right)}$, the radial tension ${\tau_m\equiv\tau_m\left(r\right)}$, and the pressure ${p_m\equiv p_m\left(r\right)}$, respectively, i.e.,
\be
T^{( m )}{}^{\mu}{}_{\nu}=\,\text{diag} (-\rho_m,\,-\tau_m,\,p_m,\,p_m)\,.
\label{EM-matter-diag}
\ee
In order to preserve the spherical symmetry of the static wormhole, we assume that these quantities depend solely on the radial coordinate $r$. Finally, the ordinary matter components can be computed through the field equations ${G_{\mu\nu}=\,T_{\mu\nu}}$, where ${T_{\mu\nu}=T_{\mu\nu}^{(B)}+T_{\mu\nu}^{(m)}}$, resulting in
\ba
\rho_m &=& -G^t{}_t-\rho_\zeta\,,\label{field1}\\
\tau_m &=& -G^r{}_r-\tau_\zeta\,,\label{field2}\\
p_m &=& G^\theta{}_\theta-p_\zeta\,,\label{field3}
\ea
respectively. The above matter equalities ensure that the solutions we will present throughout the next sections fully respect Einstein’s field equations.

\section{Wormhole solutions}\label{sec:numerics}

The Einstein equations, Eqs.~\eqref{field1}--\eqref{field3}, with the field components given in Eqs.~\eqref{eq:rho}--\eqref{eq:p}, together with the equation of motion for $\zeta$, Eq.~\eqref{eq:eom_zeta}, add up to four independent equations for seven variables, i.e., $\{\zeta,\, V,\, \Phi,\, b,\, \rho_m,\, \tau_m,\, p_m\}$. Therefore, three assumptions need to be made. Since the equation of motion, Eq.~\eqref{eq:eom_zeta}, is second order in the field $\zeta$, and only first in the potential $V$, our approach is to choose a specific form for $\zeta$ and solve it for $V$, which simplifies the computation. Accordingly, let us make the following general ans\"atze:
\be\label{ansatz}
\Phi=\Phi_0\left(\frac{r_0}{r}\right)^\alpha,\quad b=r_0\left(\frac{r_0}{r}\right)^\beta, \quad \zeta=\zeta_0\left(\frac{r_0}{r}\right)^\gamma,
\ee
where both $\gamma$ and $\alpha$ are non-negative constants and ${\beta>-1}$ in agreement with the flare-out conditions, Eq.~\eqref{def_flaring}. Here ${\Phi_0=\Phi(r_0)}$ and ${\zeta_0=\zeta(r_0)}$ are the values of the redshift function and the scalar function for the field, respectively, at the throat, $r_0$. Implementing these choices into Eq. \eqref{eq:eom_zeta}, we obtain a first-order differential equation for the potential 
\ba
\frac{\dd V}{\dd r} &=& \frac{\gamma  \zeta_0^2}{2r^3} \left(\frac{r_0}{r}\right)^{2 \gamma }\Bigg\{2 \bigg[\gamma ^2+\gamma  \left(\alpha  \Phi_0 \left(\frac{r_0}{r}\right)^{\alpha }-1\right)
	\nonumber \\
&&  +\,(\alpha -1) \alpha  \Phi_0 \left(\frac{r_0}{r}\right)^{\alpha }\bigg]\nonumber \\
&&- \left(\frac{r_0}{r}\right)^{\beta +1 } \bigg[2 \gamma ^2+\gamma  \left(\beta +2 \alpha  \Phi_0 \left(\frac{r_0}{r}\right)^{\alpha }-1\right)
	\nonumber \\
&& +\,\alpha  \Phi_0 (2 \alpha +\beta -1) \left(\frac{r_0}{r}\right)^{\alpha }\bigg]\Bigg\}\,.
\ea
The above equation can be directly integrated to give the following analytical solution
\ba\label{eq:pot}
V&=&\frac{\gamma\zeta_0^2}{2r^2}\left(\frac{r_0}{r}\right)^{2\gamma}\Bigg\{  \frac{\gamma(1-\gamma )}{\gamma +1}+2\alpha \Phi_0\left(\frac{r_0}{r}\right)^{\alpha}
	\nonumber \\
&& \times \frac{1-\alpha-\gamma}{2+\alpha+2\gamma}+\left(\frac{r_0}{r}\right)^{\beta+1}\bigg[ \frac{\gamma(2\gamma+\beta-1)}{2\gamma+\beta+3}
	\nonumber \\
&&+\, \alpha\Phi_0\left(\frac{r_0}{r}\right)^{\alpha}\frac{2\gamma+2\alpha+\beta-1}{2\gamma+\alpha+\beta+3} \bigg]\Bigg\}\,,
\ea
\begin{table}[]
\resizebox{5.4cm}{!}{\begin{tabular}{|c|c|c|c|c|c|}
\hline
 Case & $\alpha$ & $\,\,\beta\,\,$ & $\gamma$ & $\,\Phi_0\,$ & $\zeta_0$ \\ \hline
I & 0.01 & 1 & 1 & 1 & 1 \\ \hline
II & 2 & 1 & 1 & 1 & 1 \\ \hline
III & 5 & 1 & 1 & 1 & 1 \\ \hline
IV & 5 & 1 & 1 & 1 & 0.5 \\ \hline
V & 0 & 1 & 1 & 1 & 2 \\ \hline
VI & 0 & 0.02 & 0.01 & 0 & 15 \\ \hline
\end{tabular}}
\caption{\raggedright \label{tab1} Parameter values for $\alpha$, $\beta$, $\gamma$, $\Phi_0$, and $\zeta_0$ regarding a set of six different wormhole solutions for the self-interacting potential, according to Eq.~\eqref{eq:pot}.}
\end{table}
which is regular in the entire domain ${r\geqslant r_0}$ and presents the symmetry ${\zeta_0\rightarrow -\zeta_0}$. The integration constant is set to zero to ensure that $\rho_\zeta$ vanishes as ${r\rightarrow\infty}$. This also guarantees that the potential will always vanish at infinity, i.e. ${\eta=r_0/r\rightarrow0}$. To compactify the domain, we define ${\eta=r_0/r}$ which is a widely used choice when studying compact objects.

\begin{figure}[t!]
      \includegraphics[height=0.64\linewidth]{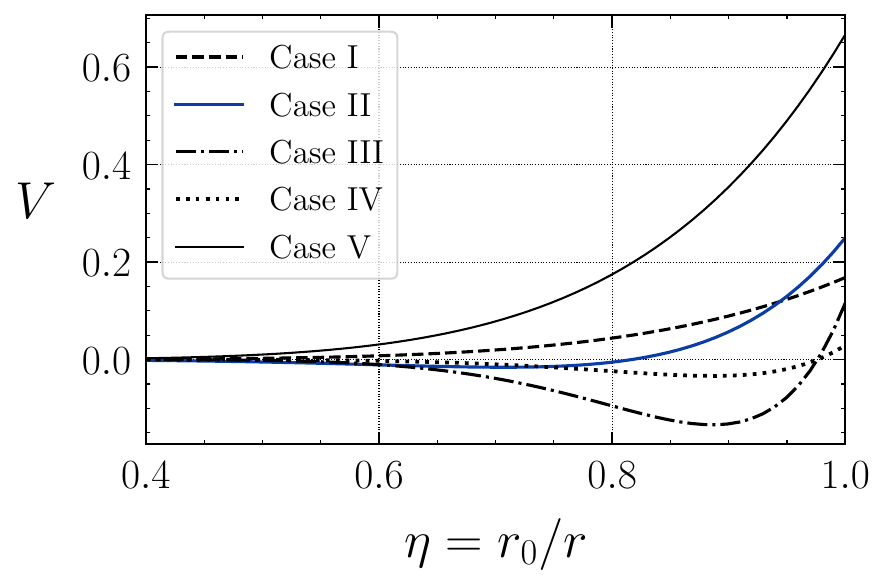}
      \captionsetup{justification=raggedright, singlelinecheck=false}
  \caption{\label{fig:pot} Self-interacting potential $V$, given by Eq.~\eqref{eq:pot}, for the specific choices of the parameters $\alpha$, $\beta$, $\gamma$, $\Phi_0$ and $\zeta_0$ as presented in Table.~\ref{tab1}. Case VI is not depicted since its potential is negligible compared to the others.}
\end{figure}

In order to study the behavior of the vector potential, we plot specific cases of Eq.~\eqref{eq:pot} in Fig.~\ref{fig:pot}, by selecting five sets of specific values for the parameters $\alpha$, $\beta$, $\gamma$, $\Phi_0$, and $\zeta_0$ as presented in Table~\ref{tab1}. Without loss of generality, we set ${r_0=1}$ in the figures of this manuscript. Furthermore, to compactify the domain, we define ${\eta=r_0/r}$. Case VI is not included in Fig.~\ref{fig:pot} since its potential $V$ is negligible when compared to the other solutions, as we show in the next subsection.

Let us first examine Cases I and V. These exhibit monotonic potentials, meaning that their first derivatives with respect to the radius $r$ are non-zero at any given radius, i.e., ${\frac{{\rm d}V}{{\rm d}r} \neq 0}$, and smoothly decay towards zero at infinity, i.e. ${\eta\rightarrow 0}$. We note that this behavior arises due to the fact that the parameter $\alpha$ is chosen to be small (${\alpha \approx 0}$) for such solutions, in contrast with other scenarios where ${\alpha \geqslant 2}$. 
The main distinction between Cases I and V lies though in the parameter $\zeta_0$. We find that larger values of $\zeta_0$ lead to a higher value of the potential at the throat, $V(r_0)$.

Now, focusing on Cases II, III and IV, a notable difference emerges. For larger values of $\alpha$, the first derivative of the potential becomes zero at a specific radius. As $\alpha$ increases from 2 to 5, the radius at which the first derivative vanishes shifts to smaller values. Comparing Cases III and IV, we observe that a higher $\zeta_0$ results in a steeper curve at the point where the first derivative becomes zero.

These observations are particularly relevant for analyzing the energy conditions in these scenarios, as they provide insights into how the shape of the potential influences the physical viability of the wormhole solutions. This discussion is presented in the following subsection.

\begin{figure*}[t!]
      \subfloat{\includegraphics[height=0.35\linewidth]{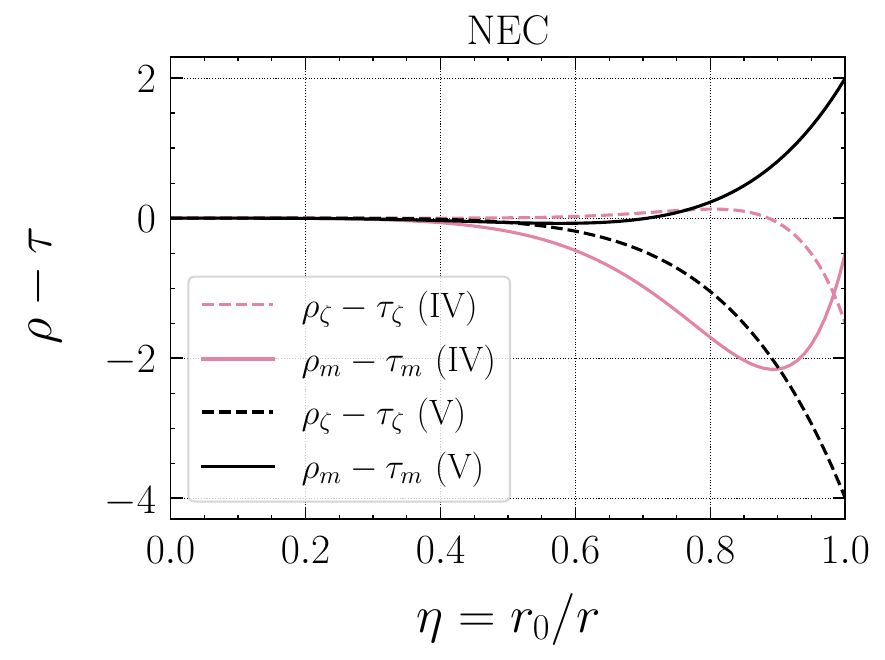}}
      \qquad
      \subfloat{\includegraphics[height=0.35\linewidth]{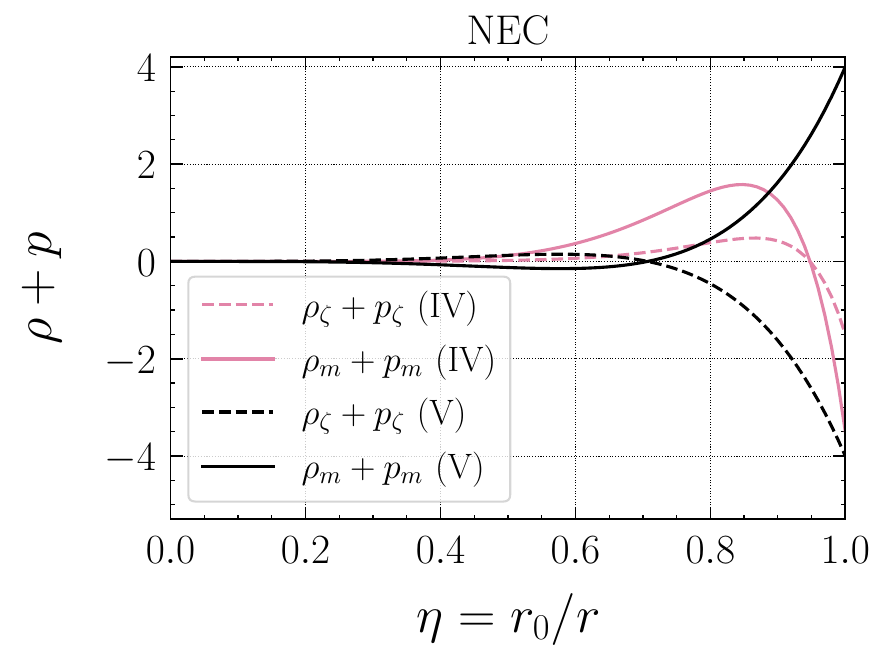}}
      \captionsetup{justification=raggedright, singlelinecheck=false}
  \caption{\label{fig:nec} Radial (left panel) and tangential (right panel) components of the NEC for the field (dashed lines) and matter (solid lines), for the Cases IV and V from Table~\ref{tab1}.}
\end{figure*}

\subsection{Energy conditions}\label{sec:EC}

Substituting the potential \eqref{eq:pot}, back into the field equations \eqref{field1}--\eqref{field3}, it is straightforward to find the physical quantities of matter, $\rho_m, \tau_m$ and $p_m$, of the wormhole solutions.

However, among the obtained solutions of the system, our focus lies mainly on those that are astrophysically relevant. That is, only on those in which ordinary matter satisfies the standard energy conditions \cite{Hawking:1973uf}. These are the null energy condition (NEC), the weak energy condition (WEC), the strong energy condition (SEC), and the dominant energy condition (DEC). For an energy-momentum tensor given by Eq. \eqref{EM-matter-diag}, these are expressed as follows \cite{Hawking:1973uf}:
\begin{eqnarray}
    \text{NEC:}&& \quad \rho_m-\tau_m \geqslant 0,\quad \rho_m+p_m \geqslant 0,
    \nonumber \\
    \text{WEC:}&& \quad \text{NEC and } \rho_m \geqslant 0,
    \nonumber \\
    \text{SEC:}&& \quad \text{NEC and } \rho_m-\tau_m+2p_m \geqslant 0,
    \nonumber \\
    \text{DEC:}&& \quad \rho_m \geqslant |\tau_m|, \quad \rho_m \geqslant |p_m|. \nonumber
\end{eqnarray}
The NEC ensures that the average energy density observed by any null observer is positive, while the WEC guarantees this for any timelike observer. The SEC maintains the attractive behavior of gravity, while the DEC guarantees that the speed of sound is smaller than the speed of light $c$.

By definition, the necessary condition for the existence of a wormhole geometrical solution is the violation of the NEC, i.e. $T_{\mu\nu}k^\mu k^\nu \geqslant 0$ where $k^\mu$ is any null vector and $T_{\mu\nu}$ is the total energy-momentum tensor. This guarantees that the flaring-out condition \eqref{def_flaring} is satisfied. Thus, since we seek solutions where the matter component $\rho_m, \tau_m$ and $p_m$ do satisfy the energy conditions, it follows that the field components $\rho_\zeta, \tau_\zeta$ and $p_\zeta$ must violate these conditions. This ensures that the massive vector field is responsible to sustain the wormhole geometry, while the ordinary matter is able to thread the wormhole without violating the classical energy conditions. Therefore, we give a special emphasis on the solutions for which only the vector is exotic.

\begin{figure*}[t!]
      \subfloat{\includegraphics[height=0.35\linewidth]{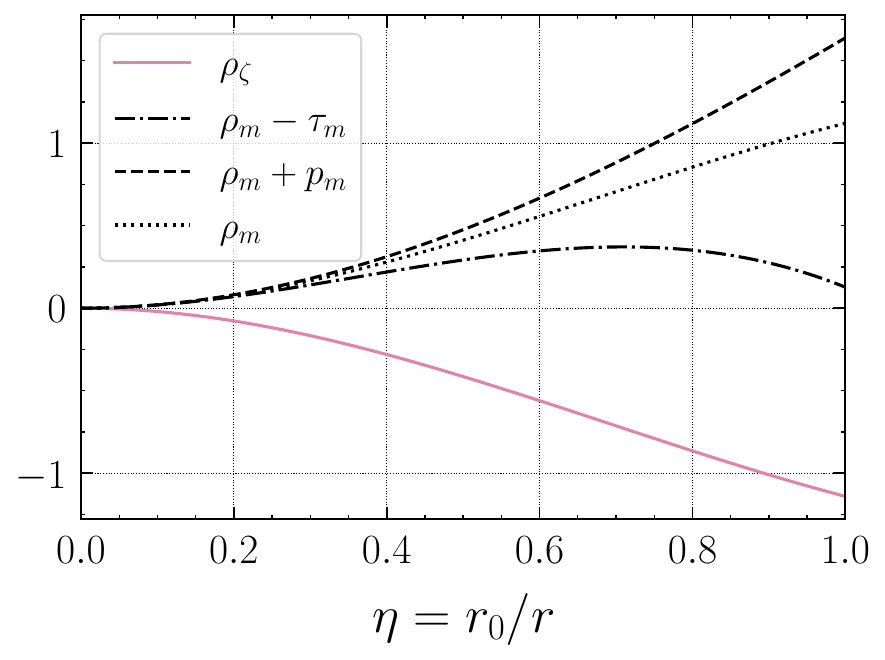}}
      \qquad
      \subfloat{\includegraphics[height=0.35\linewidth]{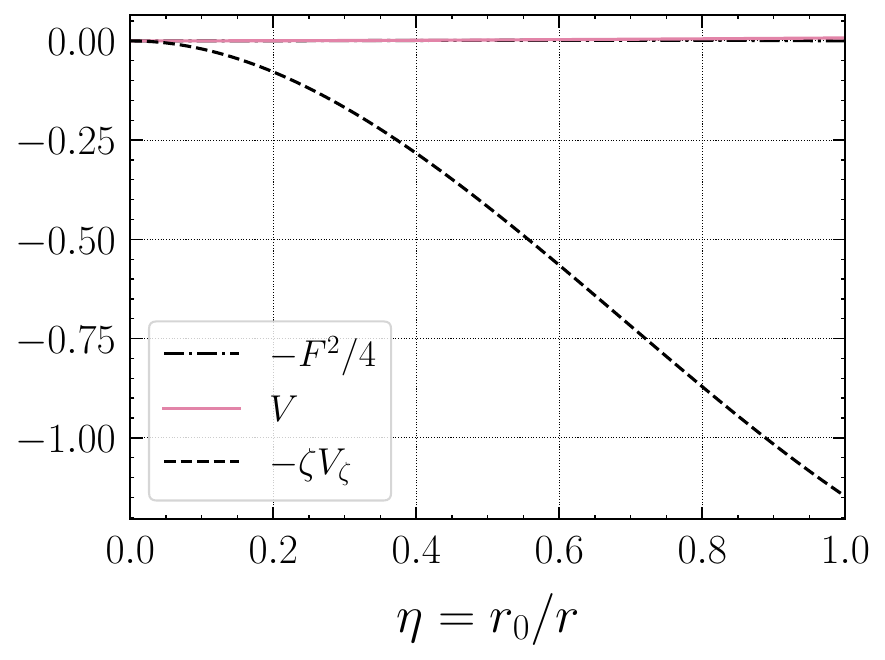}}
      \captionsetup{justification=raggedright, singlelinecheck=false}
  \caption{\label{fig:sc}\underline{\textbf{Left panel:}} Energy density and NECs for the field and matter components (see legend) corresponding to the parameters choice of Case VI according to Table~\ref{tab1}; \underline{\textbf{Right panel:}} Comparison of the terms $V$, $-F^2/4$, and $-\zeta V_{,\zeta}$ (see legend) for the solution presented in Case VI of Table~\ref{tab1}.}
\end{figure*}

At the throat, the NEC for the field components becomes
\ba
\rho_\zeta(r_0)-\tau_\zeta(r_0)&=&\rho_\zeta(r_0)+p_\zeta(r_0)
	\nonumber \\
&=& -\frac{\zeta_0^2}{2r_0^2}\left(1+\beta\right)\left(\gamma +\alpha\Phi_0\right)\,,
\ea
which tells us that for the field to violate the energy conditions at the throat, ${\rho_\zeta(r_0)-\tau_\zeta(r_0)<0}$, then ${\gamma + \alpha \Phi_0 > 0}$ is required. Through the field equations \eqref{field1}--\eqref{field3}, the above relation implies that the matter terms contributing to the NEC at the throat satisfy
\ba
\rho_m(r_0)-\tau_m(r_0) &=& \frac{1+\beta}{2r_0^2}\Big[ \zeta_0^2\left(\gamma+\alpha\Phi_0\right)-2 \Big]\,,\\
\rho_m(r_0)+p_m(r_0) &=& \rho_m(r_0)-\tau_m(r_0) 
	\nonumber \\
&& + \frac{2+\left(1+\beta\right)\left(1-\alpha\Phi_0\right)}{2r_0^2}\,.
\ea
This time, there is no straightforward relationship between the parameters which guarantees that both expressions above are non-negative. However, provided that the previously established condition $\gamma + \alpha \Phi_0 > 0$ is satisfied, we use the same cases as in Fig.~\ref{fig:pot} to identify those yielding a positive matter NEC at the throat. We observe that while multiple cases satisfy this condition at the throat, all of them eventually violate the NEC at some larger radius. 

In Fig.~\ref{fig:nec}, we compare Cases IV and V to highlight their contrasting behaviors. Case V satisfies the NEC at the throat, whereas Case IV does not. This happens due to the interplay between the values of $\alpha$ and $\zeta_0$. Among the solutions that satisfy the NEC at the throat (Cases II, III, and V), Case V is chosen here as it maintains the NEC positivity over a larger radial range. From this, we conclude that solutions with lower $\alpha$ values and higher $\zeta_0$ are more likely to result in wormhole solutions satisfying the NEC across all radii. Referring back to the results in Fig.~\ref{fig:pot}, this suggests that we should look for decreasing monotonic potentials with higher $V(r_0)$ values at the throat in order to minimize the violation of the energy conditions by the matter source.

However, we find that this alone is insufficient to ensure the non-negativity of the NEC conditions by matter in the entire domain. Therefore we are driven into an appropriate adjustment of the other parameters. By doing so, we are able to pinpoint specific parameter combinations that render ordinary matter completely non-exotic throughout the entire wormhole domain. Such is the case as the configuration presented in Case VI, illustrated in the left panel of Fig.~\ref{fig:sc}. In this specific case, ordinary matter satisfies all the classical energy conditions, from the null to the dominant, with the vector field being solely responsible for maintaining the wormhole structure. This occurs when the kinetic and potential terms, $F^2$ and $V$, are negligible compared to $\zeta V_{,\zeta}$, as depicted in the right panel of Fig.~\ref{fig:sc}. In this scenario, were the $\zeta$ function dominates, we find that ${\rho_\zeta \approx \rho_\zeta - \tau_\zeta \approx \rho_\zeta + p_\zeta \approx -\zeta V_{,\zeta}}$. This is suitable with the previous conclusion that $\zeta$ should have a large value at the throat, through the choice of $\zeta_0$. Therefore, for scenarios where $\zeta$ dominates over both the kinetic and potential terms, the vector field can sustain the wormhole geometry, ensuring that the matter components remain non-exotic throughout the entire radial domain. Recalling the previous conclusion regarding the potential, we deduce that, to obtain physically relevant wormhole solutions, the potential must not only exhibit a smooth behavior with a large value at the throat but also remain negligible in comparison with the vector field.

Finally, it is possible to find a more general analytical solution for the potential using the following shape function
\be
\label{analytical_potential}
b(r) = r_0\left(\frac{r_0}{r}\right)^\beta\exp\left[\lambda\left(1-\frac{r}{r_0}\right)\right]\,,
\ee
for which we obtain qualitatively similar results as those above, provided that the $\lambda$ parameter is fixed to any positive value. Thus, in order to lighten the content of the communication, and since there are no unfamiliar conclusions, we omit detailed results for this solution.

\section{Coupling the vector field to matter}\label{sec:couple}

\subsection{Action}\label{sec:theory2}

\begin{figure*}[t!]
      \subfloat{\includegraphics[height=0.305\linewidth]{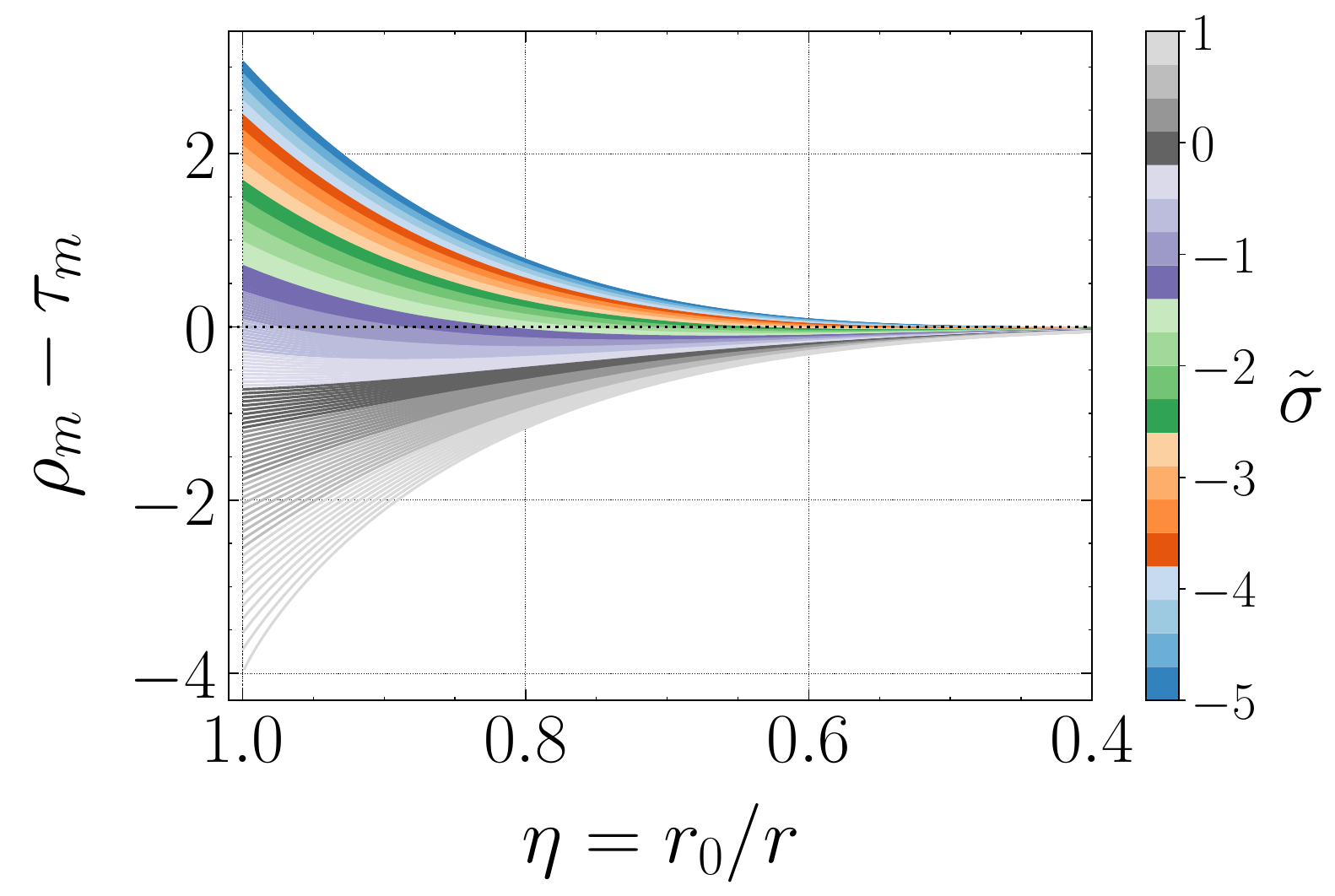}}
      \qquad
      \subfloat{\includegraphics[height=0.305\linewidth]{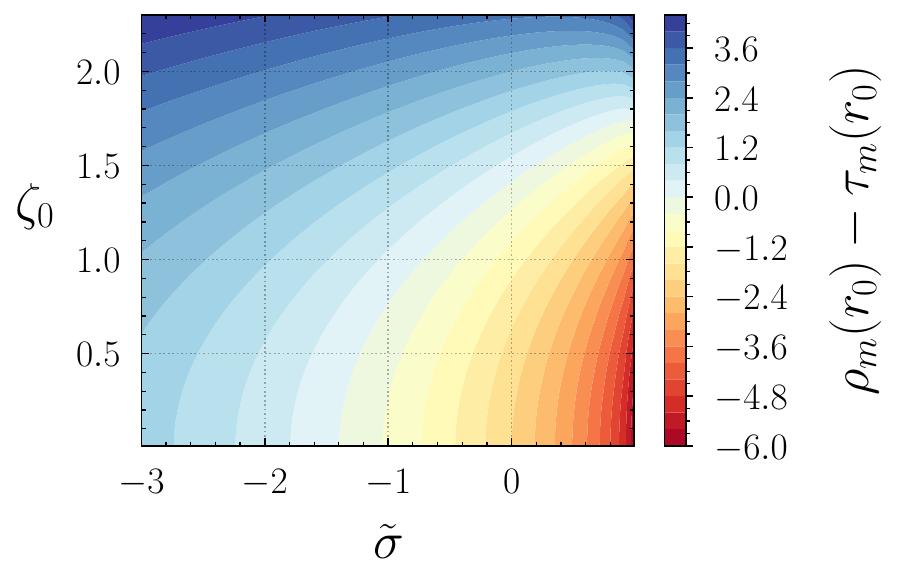}}
      \captionsetup{justification=raggedright, singlelinecheck=false}
  \caption{\label{fig:coupled} \underline{\textbf{Left panel:}} Radial matter NEC profile for the coupled case, exposed in Sec.~\ref{sec:couple}, with $\zeta_0=1$ and different values of the coupling parameter $\tilde{\sigma}$. \underline{\textbf{Right panel:}} Value of the radial component of the NEC for matter, at the throat, for different values of $\tilde{\sigma}$ and $\zeta_0$.}
\end{figure*}

In this section we consider a coupling of matter to the vector field. One natural way for this process to be realized is through the so-called conformal couplings \cite{Barros:2023pre,Bronnikov:1973fh}. Therefore, by assuming that matter follows geodesics drawn by a different metric, say $\tilde{g}_{\mu\nu}$, the action is then
\be\label{eq:action2}
\mathcal{S}=\int \dd^4x\sqrt{-g}\left[ \frac{R}{2\kappa^2}  -\frac{1}{4}F^2-V\left(B^2\right) \right]+\mathcal{S}_m\left(\psi,\tilde{g}_{\mu\nu}\right)\,.
\ee
where in the following we shall assume that this metric is related to the gravitational one through a Weyl scaling \cite{Fujii:2003pa,Faraoni:2004pi}, i.e.,
\be
\tilde{g}_{\mu\nu}=\Omega^{2}(B^2)g_{\mu\nu}\,,
\ee
with $\Omega$ being the conformal factor, here dependent on the Lorentz invariant $B^2$. The squared factor in $\Omega$ preserves the signature of the metric in both frames. The above relation naturally generates an energy flow between the vector field and matter, which becomes evident writing the conservation equations,
\ba
\nabla_\mu T^{(B)\,\mu}{}_\nu &=& -Q\,,\\
\nabla_\mu T^{(m)\,\mu}{}_\nu &=& Q\,,
\ea
where the coupling term satisfies
\be
Q = -T^{(m)}\nabla_\nu\left(\ln \Omega\right)\,.
\ee
Note that while the divergence of the individual energy-momentum tensors are not conserved, the total energy-momentum tensor is. Here $T^{(m)}$ represents the trace of the energy momentum tensor of matter, which through Eq.~\eqref{EM-matter-diag}, becomes
\be
T^{(m)} = -\rho_m-\tau_m+2p_m\,.
\ee

The equations of motion~\eqref{eq:eom_general} are now generalized to include the interaction term
\be\label{eq:motion2}
\nabla_\mu F^{\mu\nu}=2\left(\frac{\partial V}{\partial B^2}+\frac{T^{(m)}}{\Omega}\frac{\partial\Omega}{\partial B^2}\right)B^\nu\,.
\ee
Choosing a standard conformal coupling profile \cite{Barros:2023pre,Bronnikov:1973fh},
\be
\Omega \left(B^2\right) = {\rm exp}\left(-\sigma B^2\right)\,,
\ee
with $\sigma$ a constant, we have,
\ba
Q =\sigma T^{(m)}\nabla_\nu B^2 
= 2\sigma T^{(m)}B^\alpha\nabla_\nu B_\alpha\,.
\ea
Note that $\sigma$ accounts for the strength of the conformal interaction. In the coupled case, the equations of motion Eq.~\eqref{eq:motion2} get rather complicated to find an exact solution. Thus, in order to solve the system of Eqs. \eqref{field1}--\eqref{field3} and Eq.~\eqref{eq:motion2}, we start by considering the case of a constant redshift function, i.e., ${\alpha=0}$, in Eq.~\eqref{ansatz}. This transfer of energy naturally impacts the equations of motion for $\zeta$, in the following manner
\be\label{eq:eom2_zeta}
\zeta''\left(1-\frac{b}{r}\right)+\frac{\zeta'}{r}\left( 2-\frac{3}{2}\frac{b}{r}-\frac{b'}{2} \right)+V_{,\zeta}=-2\sigma T^{(m)}\zeta\,.
\ee
The right-hand side term in the above equation stems from the coupling and has a significant impact on the overall geometrical solutions, as we show in the next subsection.

\subsection{Wormhole solutions}\label{subsec:solutions}

We can now use the Einstein equations~\eqref{field1}--\eqref{field3} to substitute the matter quantities in Eq. \eqref{eq:eom2_zeta}. Solving the equation for the simplified case ${\beta=\gamma=1}$, one obtains
\be
\frac{\dd V}{\dd r} = -r_0^2\zeta_0^2\frac{r_0^2-4\sigma\left(r_0^2+2r^4V\right)}{r^5(r^2-2\sigma r_0^2\zeta_0^2)}\,,
\ee
which possesses the following analytical solution
\ba\label{eq:V_coupled}
V(\eta) &=& \frac{2\tilde{\sigma}-\zeta_0^2}{4\tilde{\sigma}^3r_0^2}
\Big[ \tilde{\sigma}\eta^2\left(2-3\tilde{\sigma}\eta^2\right)
	\nonumber \\
&& -2\left(1-\tilde{\sigma}\eta^2\right)^2\ln\left(1-\tilde{\sigma}\eta^2\right) \Big]  \,,
\ea
where we redefine the coupling strength as ${\tilde{\sigma}=2\zeta_0^2\sigma}$, for nonzero values of $\tilde{\sigma}$ and $r_0$. We impose ${\eta^2<\tilde{\sigma}^{-1}}$ which gives an upper bound $\tilde{\sigma}<1$. Note that the above solution is again symmetric under ${\zeta_0\rightarrow -\zeta_0}$, since it only depends on the square of this parameter. This solution describes a wormhole with a constant redshift function, and an inverse radial function for the field and shape function, i.e., ${b\propto\zeta\propto 1/r}$. Consistently, the limit where the coupling vanishes, i.e., ${\sigma\rightarrow 0}$, one recovers Eq.~\eqref{eq:pot} with ${\left\{\alpha,\beta,\gamma\right\}=\left\{0,1,1\right\}}$,
\be
\lim_{\sigma\rightarrow 0}V(\eta)=\frac{\zeta_0^2\eta^6}{6r_0^2}\,.
\ee

The radial NEC profile for matter generated by solution \eqref{eq:V_coupled} is thus shown in the left panel of Fig.~\ref{fig:coupled}, where we set $\zeta_0=1$ and consider different values of the coupling parameter $\tilde{\sigma}$. We only plot the radial component since the tangential component exhibits the same behavior and
\be
\rho_m-\tau_m<\rho_m+p_m\,.
\ee
Upon careful examination, it becomes clear that when the parameter $\tilde{\sigma}$ is close to zero (noninteracting case), the matter distribution violates the NEC, specifically through its radial component. However, stronger couplings, through negative values of $\tilde{\sigma}$, can modify this behavior, rendering the matter non-exotic near the throat. This effect can be achieved similarly with higher values of $\zeta_0$. In such scenarios, the matter distribution can be adjusted to satisfy the NEC close to the throat, thereby avoiding its violation.  The right panel of Fig.~\ref{fig:coupled} illustrates this feature, where one can observe that larger values of either $|\tilde{\sigma}|$ or $\zeta_0^2$ can indeed ensure non-exotic matter at the throat. Nevertheless, note that this compliance is limited to the vicinity of the throat. 
This becomes apparent when observing that, although the NEC profiles decay toward zero as ${r \to \infty}$ (or ${\eta \to 0}$), they do so through negative values. To demonstrate this, one can compute the dominant term in the limit ${r \to \infty}$, i.e.,
\be
\left(\rho_m-\tau_m\right)|_{\eta=0}\approx-\frac{2 \eta ^4}{r_0^2}+\mathcal{O}\left(\eta ^6\right)\,,
\ee
which is negative. This signals that the matter components inevitably violate the NEC at some radial distance $r$ far from the throat.

Finally, we have also explored solutions with ${\alpha = 2}$, which introduce $\Phi_0$ as a free parameter into the solution for the potential, in contrast to the previous case described in Eq.~\eqref{eq:V_coupled}. We find qualitatively similar results to those discussed earlier, provided that the parameter $\Phi_0$ is fixed to any non-positive value. As no additional insights are gained from these solutions, we omit showing the detailed results.

\section{Conclusions}\label{sec:conclusions}

This study has focused on the dynamics of a massive vector field, \textbf{B}, either minimally or non-minimally coupled to Einstein gravity, with the aim of investigating its potential role in supporting wormhole spacetimes.

Our results indicate that massive one-forms, when minimally coupled to matter and despite the challenges associated with their stability, present a promising mechanism for satisfying the energy conditions essential for wormhole geometries. Additionally, we examined the interactions between the massive vector field and matter fields through conformal couplings. This second analysis allowed us to explore the effects of these couplings on the energy conditions, shedding light on the physical plausibility of wormhole solutions. Specifically, we have concluded that it is possible for matter to thread the wormhole without violating any energy condition in the case of a non-interacting vector field supporting the wormhole. On the other hand, when the massive vector is coupled to ordinary matter, although a strong interaction is able to make matter non-exotic close to the throat, it will always, albeit minimally, violate the null energy condition at some distance to the throat. Nonetheless, this specific radius can be put close to infinity. These results not only enhance our understanding of wormhole physics but also provide valuable insights into the broader role of vector fields in the context of modified theories of gravity and other exotic geometries. 

Ultimately, this work represents a contribution to the expanding field of wormhole physics, particularly by highlighting the potential of massive one-forms as a viable framework for theoretically constructing wormhole solutions. While significant challenges remain, especially concerning the stability and alignment of these theories with fundamental physical principles, our results emphasize the need for continued exploration into the dynamics of massive vector fields. Such efforts are crucial for advancing both theoretical understanding and phenomenological applications in modern gravitational theories.

\acknowledgments

The authors thank Tiago Barreiro and Jo\~ao Lu\'is Rosa for enlightening discussions.
NG, BJB and FSNL acknowledge support from the Funda\c{c}\~{a}o para a Ci\^{e}ncia e a Tecnologia (FCT) through the project BEYLA: BEYond LAmbda with reference number PTDC/FIS-AST/0054/2021, and through the research grants UIDB/04434/2020, UIDP/04434/2020.
FSNL acknowledges support from the Funda\c{c}\~{a}o para a Ci\^{e}ncia e a Tecnologia (FCT) Scientific Employment Stimulus contract with reference CEECINST/00032/2018.
AdlCD acknowledges support from BG20/00236 action (MCINU, Spain), NRF Grant CSUR23042798041 (South Africa), CSIC Grant COOPB23096 (Spain), Project SA097P24 funded by Junta de Castilla y Le\'{o}n (Spain), Grants PID2021-122938NB-I00 and  PID2022-137003NB-I00 funded by MCIN/AEI/10.13039/501100011033 and by \textit{ERDF A way of making Europe}.

\bibliography{bib}

\end{document}